\documentstyle[preprint,aps]{revtex}
\input BoxedEPS.tex
\SetOzTeXEPSFSpecial
\HideDisplacementBoxes

\begin{document}


%
\title{Large oxygen-isotope effect in
Sr$_{0.4}$K$_{0.6}$BiO$_{3}$: Evidence for phonon-mediated
superconductivity}
\author{Guo-meng Zhao$^{(1)}$, K.Conder$^{(2)}$, M.Angst$^{(3)}$, S.
M. Kazakov$^{(3)}$,
J. Karpinski$^{(3)}$, M. Maciejewski $^{(4)}$, C. Bougerol$^{(5)}$,
J. S. Pshirkov$^{(6)}$, E. V. Antipov$^{(6)}$, }

\address {
$^{(1)}$ Physik-Institut der Universit\"at Z\"urich, CH-8057
Z\"urich, Switzerland\\
$^{(2)}$ Laboratorium f\"ur Neutronenstreuung ETH and PSI, 5232 Villigen PSI,
Switzerland\\
$^{(3)}$ Laboratorium f\"ur Festk\"orperphysik, ETH
Z\"urich, CH-8093 Z\"urich, Switzerland\\
$^{(4)}$ Laboratorium f\"ur Technische Chemie, ETH
Z\"urich, CH-8093 Z\"urich, Switzerland\\
$^{(5)}$ Laboratoire de Cristallographie, CNRS-UJF, BP 166, 38042
Grenoble
cedex 9, France\\
$^{(6)}$ Department of Chemistry, Moscow State University, Moscow
119899,
Russia\\}

\maketitle
\widetext
%
\begin{abstract}
Oxygen-isotope effect has been investigated in a recently
discovered superconductor Sr$_{0.4}$K$_{0.6}$BiO$_{3}$. This compound has a
distorted perovskite structure and becomes superconducting at about 12
K. Upon replacing $^{16}$O with
$^{18}$O by 60-80$\%$, the $T_{c}$ of the sample is shifted down by
0.32-0.50 K, corresponding to an isotope exponent of $\alpha_{O}$ =
0.40(5). This
isotope exponent is very close to that for a similar bismuthate
superconductor Ba$_{1-x}$K$_{x}$BiO$_{3}$
with $T_{c}$ = 30 K. The very distinctive doping and $T_{c}$
dependencies of $\alpha_{O}$ observed in bismuthates and cuprates suggest that
bismuthates should belong to conventional
phonon-mediated superconductors while cuprates might be unconventional
supercondutors.

\end{abstract}
\newpage
\narrowtext
The discovery of high-temperature superconductivity near 30 K in
the nonmagnetic cubic perovskite oxide Ba$_{1-x}$K$_{x}$BiO$_{3}$
(BKBO)
\cite{Mattheiss,Cava} raises an interesting question of whether the
layered structure and strong antiferromagnetic correlation in
cuprates are essential for high-temperature superconductivity. In
order to answer this question, it is important to find some common
and distinct features in both systems. The band-structure
calculations \cite{Mat} suggest that the bare density of states at Fermi level
in BKBO is at least 3 times smaller than that in cuprates.  Since tunneling
and extensive
oxygen-isotope experiments on BKBO \cite{Huang,Batlogg,Hinks,Kondoh,Zhao}
seem to indicate that this material is a conventional phonon-mediated
superconductor
with an electron-phonon coupling constant $\lambda$ $\sim$ 1, one might argue
that 100 K superconductivity in cuprates could be understood even
within the conventional theory with $\lambda$ $\sim$ 3. If this were the case,
the isotope effects in both cuprates and bismuthates would be
similar. As a matter of fact, the oxygen-isotope exponent $\alpha_{O}$
in BKBO has a maximum at optimal doping where $T_{c}$ is the highest
\cite{Zhao}, while $\alpha_{O}$ in optimally-doped cuprates is
the smallest \cite{Crawford1,Franck,Morris,ZhaoNature}. Moreover, the
isovalent substitution of Ca for Sr in
the single-layer La$_{2-x}$Sr$_{x}$CuO$_{4}$ system leads to a large
decrease in
$T_{c}$, and to a large increase in $\alpha_{O}$ \cite{Crawford2}. Similarly,
the isovalent substitution of Sr for Ba in
YBa$_{2}$Cu$_{3}$O$_{7}$ gives rise to a strong suppression of
superconductivity from 93 K to 60 K \cite{Okai}, and $\alpha_{O}$ also
increases with decreasing $T_{c}$ \cite{ZhaoPhysicaC}. If the
pairing mechanism in bismuthates and cuprates were the same, this
unusual $T_{c}$ dependence of $\alpha_{O}$ would also exist in
bismuthates.

The isovalent substitution of Sr for Ba in Ba$_{1-x}$K$_{x}$BiO$_{3}$
cannot be realized using conventional solid-state  reaction. Until recently,
a new family of bismuth-oxide-based superconductor
Sr$_{1-x}$K$_{x}$BiO$_{3}$ (SKBO) has been synthesized
by a high pressure
technique \cite{Kazakov}. This material has a distorted perovskite
structure and
exhibits a superconductivity at about 12 K, which is much lower than
that in Ba$_{1-x}$K$_{x}$BiO$_{3}$. It appears that the
isovalent substitution effects on $T_{c}$ in
both YBa$_{2}$Cu$_{3}$O$_{7}$ and Ba$_{1-x}$K$_{x}$BiO$_{3}$ are
quite similar. Now the important question is whether $\alpha_{O}$ in
optimally-doped Sr$_{1-x}$K$_{x}$BiO$_{3}$ remains the same or
increases substantially compared with the exponent in optimally-doped
Ba$_{1-x}$K$_{x}$BiO$_{3}$. The clarification of this issue
will provide important insight into the
pairing mechanism of high temperature superconductivity in both
bismuthates and cuprates.

Here we report the oxygen-isotope effect
in optimally-doped Sr$_{1-x}$K$_{x}$BiO$_{3}$ ($x$ = 0.6) with
$T_{c}$ = 12 K.  We
found a large oxygen-isotope
exponent $\alpha_{O}$ = 0.40(5), which is the same as
for optimally doped Ba$_{1-x}$K$_{x}$BiO$_{3}$. The present
isotope experiments clearly demonstrate that
the conventional phonon-mediated mechanism is responsible for the
superconductivity in bismuthates, and that the pairing mechanism in
cuprates is unconventional.

The samples were synthesized by a high pressure technique \cite{Kazakov}.
First,
the sample of Sr$_{2}$Bi$_{2}$O$_{5}$ was prepared by
conventional solid state reaction using SrCO$_{3}$ and
Bi$_{2}$O$_{3}$. The
powders were mixed, ground thoroughly, and then fired in air at
700, 800, and 850 $^{\circ}$C for 100 hours with several intermediate
grindings.
Stoichiometric amounts of Sr$_{2}$Bi$_{2}$O$_{5}$, KO$_{2}$ and
Bi$_{2}$O$_{3}$
were
mixed in a dry box (filled with argon) and packed into the gold
capsules. The high pressure synthesis was carried out in a Belt-type
apparatus at 2 GPa and 700 $^{\circ}$C for 0.5 h.
Oxygen isotope exchange was performed in a closed system and at 1 bar
oxygen pressure. The
$^{18}$O
sample of pair I/pair II was prepared by annealing the
powder sample at 350/375 $^{\circ}$C for 100 h in an
$^{18}$O$_{2}$ atmosphere (ISOTEC Inc. 97$\%$ $^{18}$O$_{2}$).
The $^{16}$O control samples were
annealed in the same condition as the $^{18}$O samples, but
in an $^{16}$O$_{2}$ atmosphere.

The oxygen-isotope enrichments of the present samples cannot be
reliably determined by the conventional method commonly used in the
isotope experiments on the cuprates \cite{Conder}.
This is because the $^{18}$O sample
can be decomposed during the back-exchange with the $^{16}$O isotope in
the thermobalance. We, therefore,  proposed
a new method to determine the isotope
content precisely. This method is based on the mass spectrometric
determination of the produced water when the sample is
reduced by hydrogen.
The reduction was carried out on a Netzsch STA 405
thermoanalyzer connected to a Balzers QMG 420 quadrupole mass
spectrometer (MS) by a heated (at about 200 $^{\circ}$C) capillary.
The mixed gas (20 vol$\%$ hydrogen and 80 vol$\%$ helium) was flowing
through the thermoanalyzer with a
rate of about 50 ml/min. The heating rate was 10 $^{\circ}$C/min.
The integrated intensities of the m/z = 20 (H$_{2}$$^{18}$O) and
m/z = 18 (H$_{2}$$^{16}$O)
signals were used
for the
determination of the ratio of evolved H$_{2}$$^{16}$O and
H$_{2}$$^{18}$O. The decomposition of NaHCO$_{3}$ into H$_{2}$O and
CO$_{2}$ was used to calibrate the contribution of
m/z = 20 (H$_{2}$$^{18}$O) signal from the
$^{16}$O sample. The details of the calibration were given in
Ref.~\cite{Mac1,Mac2}.

Fig.~1 shows the intensities of the
mass spectrometric signals of m/z = 20 (H$_{2}$$^{18}$O)
and m/z = 18 (H$_{2}$$^{16}$O)
for the $^{16}$O
sample (dash
lines) and the $^{18}$O sample (solid lines) of pair I. The reduction of
the
oxides occurs in the temperature range 300-600 $^{\circ}$C with two
steps
at about 430 and
530 $^{\circ}$C for the $^{16}$O sample, and at
about 440 and 520 $^{\circ}$C for the $^{18}$O sample.
From the integrated intensities of m/z = 20 (H$_{2}$$^{18}$O)
and m/z = 18 (H$_{2}$$^{16}$O)
signals, we calculated that the $^{18}$O sample of pair I
contains 60(2)$\%$ $^{18}$O isotope. The $^{18}$O sample of pair II
contains $\sim$80$\%$ $^{18}$O isotope.

The isotope-exchanged samples were characterized by x-ray diffraction
technique
using a STADI-P diffractometer equipped with a mini-PSD detector and
a Ge monochromator on the
primary beam. The diffraction patterns were recorded in the 2$\theta$
=
10-90$^{\circ}$ range in a transmission mode by rotating the sample.
Fig.~2 shows the x-ray diffraction pattern of the $^{18}$O sample
of pair I. All peaks can be indexed on the basis of
a distorted
perovskite cell with lattice parameters $a\simeq b\simeq
\sqrt{2}a_{p}$ and  $c \simeq 2a_{p}$, where $a_{p}$
refers to the ideal cubic perovskite cell ($a_{p}\sim$ 4 \AA). No
impurity phase can be seen from the spectrum. On the other hand,
the samples (pair II) annealed at 375 $^{\circ}$C contain small
amount of impurity phase, namely,  SrBi$_{2}$O$_{4}$ (about 5$\%$).

For magnetic measurements, we pressed the samples into
pellets with a diameter of 3 mm and sealed them in respective quartz
tubes. Magnetization of the samples was measured using a SQUID magnetometer
in a magnetic field of 1 mT.
Measurements were carried out in field cooled condition, and the data
were corrected upon warming.

Fig. 3 shows the field-cooled susceptibility of the
$^{16}$O and $^{18}$O samples of pair I over 2-15 K. It is apparent
that $T_{c}$ for the
$^{18}$O
is lower than for the $^{16}$O sample. The $T_{c}$ (diamagnetic onset
temperature) of the $^{16}$O
sample is about 11.5 K, in good agreement with the value reported in
Ref.~\cite{Kazakov}. Furthermore, the low-temperature
susceptibility for the $^{18}$O sample is lower than for the $^{16}$O
sample by about 7(2)$\%$. We are not sure that this is a real effect.
The isotope back-exchange is required to clarify this, but we were
unable to do the back-exchange due to the fact that a further
increase of the  annealing time  will lead to a decomposition of the sample.

In Fig.~4, we plot the normalized magnetizations for the
$^{16}$O and $^{18}$O samples of pair I and pair II. One can see that
the transition
curves are parallel for the
$^{16}$O and $^{18}$O samples of pair I (Fig.~4a), while this is
not the case for pair II (Fig.~4b).  In order to eliminate effects due
to possible differences in demagnetization factor, particle size, and
superconducting fraction of two isotope samples, the isotope shifts of
$T_{c}$ are determined from the linear portion of the magnetization
data extended to the base line at zero magnetization. The
isotope shift is 0.32(2) K for pair I, and 0.50(4) K for pair II.
The isotope effect exponent $\alpha_{O}$ was calculated from the
definition $\alpha_{O} = - d\ln T_{c}/d\ln M_{O}$,
where $M_{O}$ is the atomic mass of the oxygen-isotopes corrected for the
incomplete exchange. We find that $\alpha_{O}$ = 0.37(3) for pair I,
and 0.42(4) for pair II. Both the values are very close to the
exponent found for Ba$_{1-x}$K$_{x}$BiO$_{3}$ with the highest
$T_{c}$ \cite{Hinks,Zhao}.

The most striking feature we found for the bismuthate superconductors
is that the oxygen-isotope exponent reaches a maximum at optimal doping, and
that the maximum value is close to 0.5,
independent of $T_{c}$. This is in sharp contrast to the isotope
effects in cuprates where $\alpha_{O}$ becomes very small at optimal
doping. Moreover, the
isovalent substitution of Ca for Sr in
La$_{2-x}$Sr$_{x}$CuO$_{4}$ system leads to a large decrease in
$T_{c}$, and to a large increase in $\alpha_{O}$ \cite{Crawford2}.
Similar result has been found for the isovalent substitution of Sr for Ba in
YBa$_{2}$Cu$_{3}$O$_{7}$ \cite{ZhaoPhysicaC}. The $T_{c}$
dependence of $\alpha_{O}$ for optimally-doped bismuthates and
cuprates are very different as shown in Fig.~5. It is remarkable
that $\alpha_{O}$ in
bismuthates is nearly independent of $T_{c}$, whereas $\alpha_{O}$ in
cuprates increases linearly with decreasing $T_{c}$. The very distinctive
isotope effects observed in the bismuthates and cuprates strongly suggest that
the microscopic
superconducting mechanisms in the two systems should be different.

It is known that the isotope exponent in conventional phonon-mediated
superconductors is close to 0.5, and nearly independent of $T_{c}$.
The fact that the isotope exponent in optimally-doped bismuthates is
also close to 0.5 and independent of $T_{c}$ (see Fig.~5) provides
strong evidence for the conventional phonon-mediated pairing mechanism
in this system. The very different isotope effects observed in
cuprates indicate that the microscopic pairing mechanism in this system
should be unconventional. The unconventional pairing mechanism in cuprates
should be related to the unique features such as two-dimensionality,
antiferromagnetic fluctuations and polaronic effects. Any feasible
theories for high-temperature superconductivity in cuprates must be
able to explain the unusual doping and $T_{c}$ dependencies of the
oxygen-isotope exponent.

In summary, oxygen-isotope effect has been investigated in a recently
discovered superconductor Sr$_{1-x}$K$_{x}$BiO$_{3}$. This material
can be synthesized by a high pressure technique.  At an
optimal doping of $x$ = 0.6, the compound exhibits superconductivity
at about 12
K. Upon replacing $^{16}$O with
$^{18}$O by 60-80$\%$, the $T_{c}$ of the sample is shifted down by
0.32-0.50 K, corresponding to an isotope exponent of $\alpha_{O}$ =
0.40(5). This
isotope exponent is very close to that for a similar bismuthate
superconductor Ba$_{1-x}$K$_{x}$BiO$_{3}$
with $T_{c}$ = 30 K. The present results clearly suggest that
bismuthates should belong to conventional
phonon-mediated superconductors while cuprates might be unconventional
supercondutors.

Acknowledgments.
This work has been performed in the framework of SNSF grant (7SUPJ
048713) and partially supported by PICS-RFBR (98-03-22007) grant.

\bibliographystyle{prsty}

\begin{thebibliography}{10}
\bibitem{Mattheiss}L.R. Mattheiss, E.M. Gyorgy, D.W. Johnson, Phys.
Rev. B
\textbf{37}, 3745 (1988).
\bibitem{Cava}R.J. Cava, B. Batlogg, J.J. Krajewski, R. Farrow, L.W.
Rupp Jr.,
A.E. White, K. Short, W.F. Peck, T. Kometani, Nature (London)
\textbf{332},
814 (1988).
\bibitem{Mat} L. F. Matteiss and D. R. Hamann, Phys.
Rev. Lett. \textbf{58}, 1028 (1987); L. F. Matteiss and D. R. Hamann, Phys.
Rev. Lett. \textbf{60}, 2681 (1988).
\bibitem{Huang} Q. Huang, J. F. Zasadzinski, N. Tralshawala, K. E.
Gray, D.G. Hinks, J. L. Peng, and R. L. Greene,
Nature (London) \textbf{347}, 369 (1990).
\bibitem{Batlogg} B. Batlogg, R.J. Cava, L.W. Rupp Jr., A.M. Mujsce,
J.J. Krajewski,
J.P. Remeika, W.F. Peck Jr., A.S. Cooper, and G.P. Espinosa, Phys.
Rev. Lett. \textbf{61}, 1670 (1988).
\bibitem{Hinks} D.G. Hinks, D.R. Richards, B. Dabrowski, D.T. Marx,
A.W. Mitchell,
Nature (London) \textbf{335}, 419 (1988).
\bibitem{Kondoh} S. Kondoh, M. Sera, Y. Ando, and M. Sato, Physica C
\textbf{157}, 469
(1989).
\bibitem{Zhao} G. M. Zhao and D. E. Morris Phys. Rev. B \textbf{51},
12 848 (1995).
\bibitem{Crawford1}M. K. Crawford, M.N. Kunchur, W. E. Farneth, E. M.
McCarron III, and S. J. Poon, ~Phys. Rev.
B \textbf{41}, 282 (1990).
\bibitem{Franck}J. P. Franck, S. Harker, and J. H. Brewer,~Phys. Rev.
Lett. \textbf{71}, 283 (1993).
\bibitem{Morris}H. J.  Bornemann and D. E. Morris, Phys.
Rev. B \textbf{44}, 5322 (1991).
\bibitem{ZhaoNature}G. M. Zhao,  M. B. Hunt, H. Keller, K. A.
M\"uller,
Nature (London) \textbf{385}, 236 (1997).
\bibitem{Crawford2}M. K. Crawford, W. E. Farneth, R. Miao, R. L.
Harlow, C. C. Torardi, and E. M.
McCarron III, ~Physica.
C \textbf{185-189}, 1345 (1991).
\bibitem{Okai}B. Okai, Jpn. J.  Appl. Phys. \textbf{29},
L2180 (1990).
\bibitem{ZhaoPhysicaC}G. M. Zhao, A. P. B. Sinha, and D. E. Morris, ~Physica.
C \textbf{297}, 23 (1998).
\bibitem{Kazakov} S.M. Kazakov, C. Chaillout, P. Bordet, J.J.
Capponi, M.
Nunez-Regueiro, A. Rysak, J.L. Tholence, P.G. Radaelli, S.N. Putilin,
E.V. Antipov, Nature (London) \textbf{390}, 148 (1997).
\bibitem{Conder}K. Conder, Ch. Kr\"uger, E. Kaldis, G. Burri and L.
Rinderer, Mat.
Res. Bull. \textbf{30}, 491 (1995).
\bibitem{Mac1} M. Maciejewski, and A. Baiker, Thermochim. Acta
\textbf{295}, 95 (1997).
\bibitem{Mac2} M. Maciejewski, C.A. M\"uller, R. Tschan, W.-D. Emmerich,
and A.
Baiker, Thermochim. Acta \textbf{295}, 167 (1997).



\end{thebibliography}

\newpage
\begin{figure}[b]
\vspace{2cm}
    \ForceWidth{14cm}
	\centerline{\BoxedEPSF{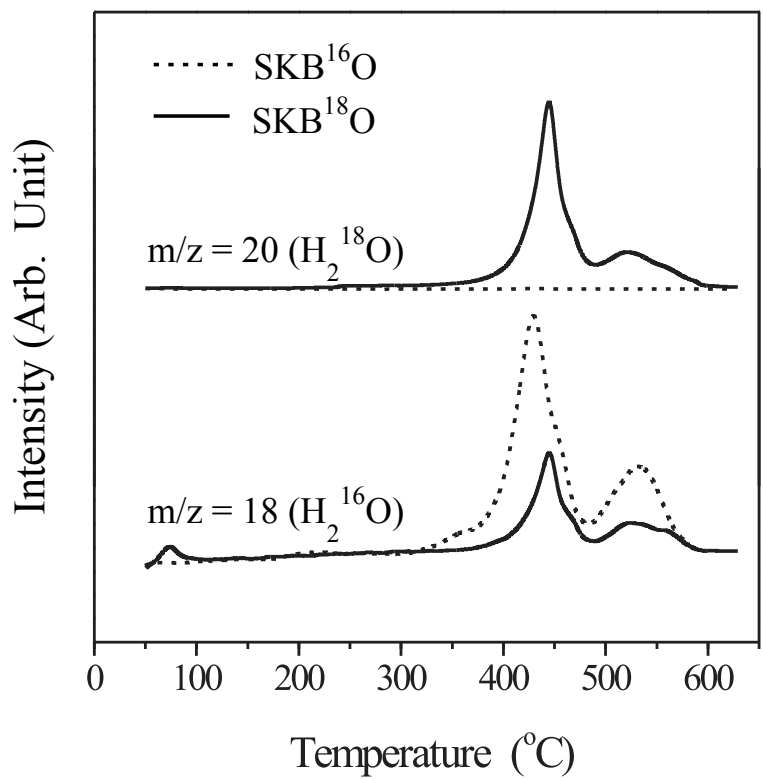}}
	\vspace{-2cm}
	\caption[~]{The intensities of the
mass spectrometric signals of m/z = 20 (H$_{2}$$^{18}$O) and m/z = 18
(H$_{2}$$^{16}$O)
for the $^{16}$O
sample (dash lines) and the $^{18}$O sample (solid lines) of pair I. From
the integrated intensities of the signals, we find that the $^{18}$O sample
contains 60(2)$\%$ $^{18}$O
isotope.}
	\protect\label{Fig.1}
\end{figure}
\newpage
~\\
\begin{figure}[b]
\vspace{2cm}
    \ForceWidth{17cm}
	\centerline{\BoxedEPSF{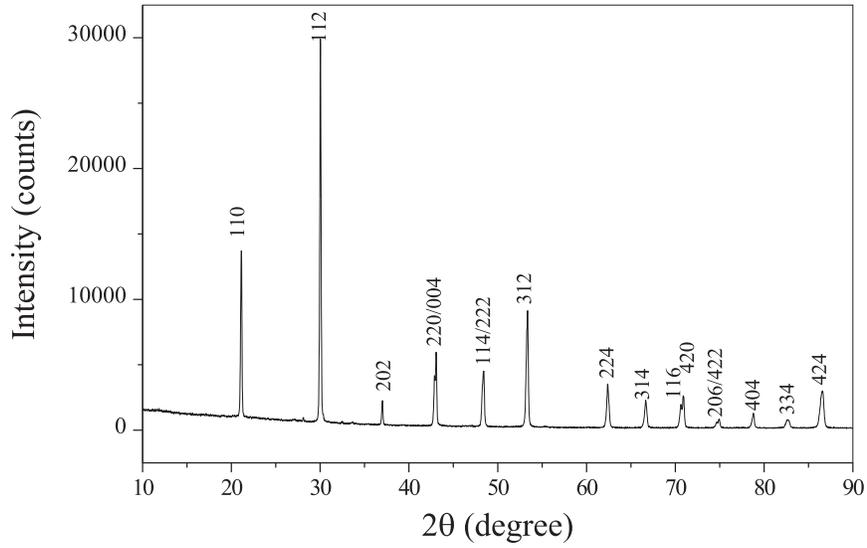}}
	\vspace{-4cm}
	\caption[~]{X-ray diffraction pattern of the $^{18}$O sample
of pair I. All peaks can be indexed on the basis of
a distorted
perovskite cell with lattice parameters $a\simeq b\simeq
\sqrt{2}a_{p}$ and $c \simeq 2a_{p}$, where $a_{p}$
refers to the ideal cubic perovskite cell ($a_{p}\sim$ 4 \AA). No
impurity phase can be seen from the spectrum.}
	\protect\label{Fig.2}
\end{figure}

\newpage
~\\
~\\
\begin{figure}[b]
\vspace{4cm}
    \ForceWidth{12cm}
	\centerline{\BoxedEPSF{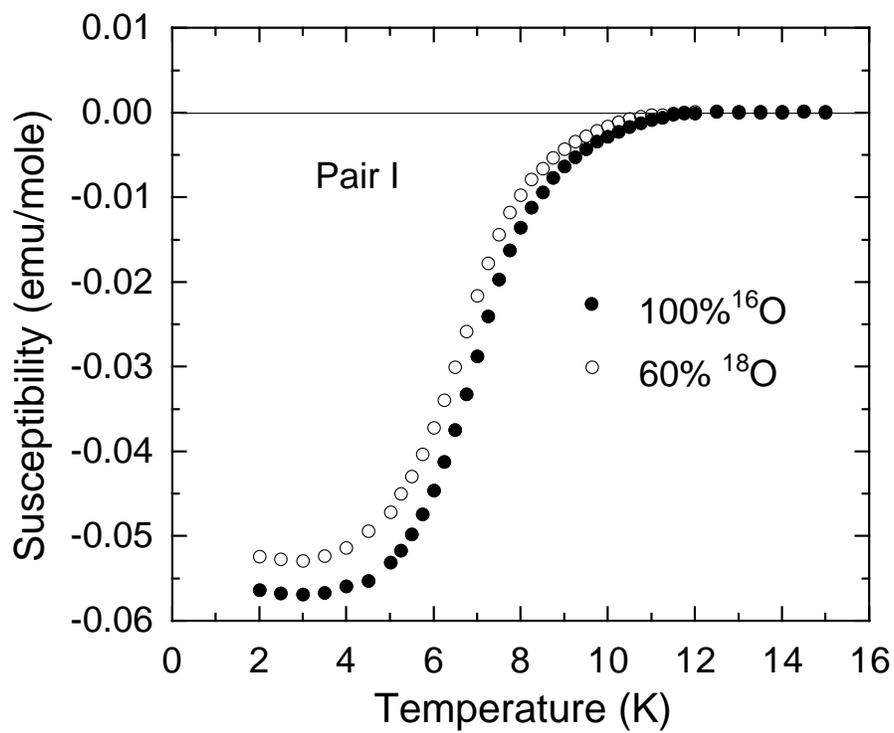}}
	\vspace{1cm}
	\caption[~]{The field-cooled susceptibility of the
$^{16}$O and $^{18}$O samples of pair I over 2-15 K.
The $T_{c}$ (diamagnetic onset
temperature) of the $^{16}$O
sample is about 11.5 K.}
	\protect\label{Fig.3}
\end{figure}
\newpage
~\\
\begin{figure}[b]
    \ForceWidth{10cm}
	\centerline{\BoxedEPSF{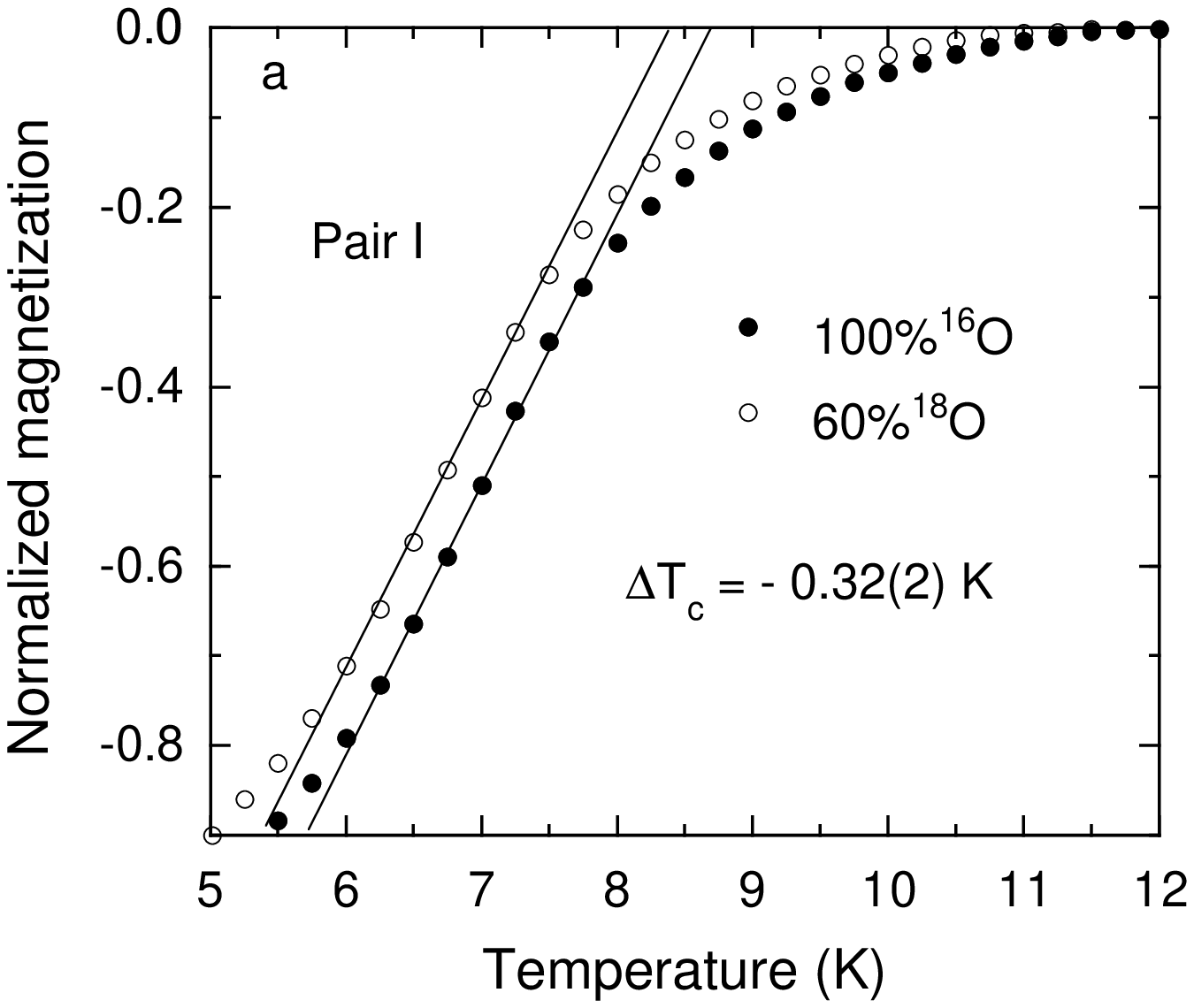}}
	\ForceWidth{10cm}
	\centerline{\BoxedEPSF{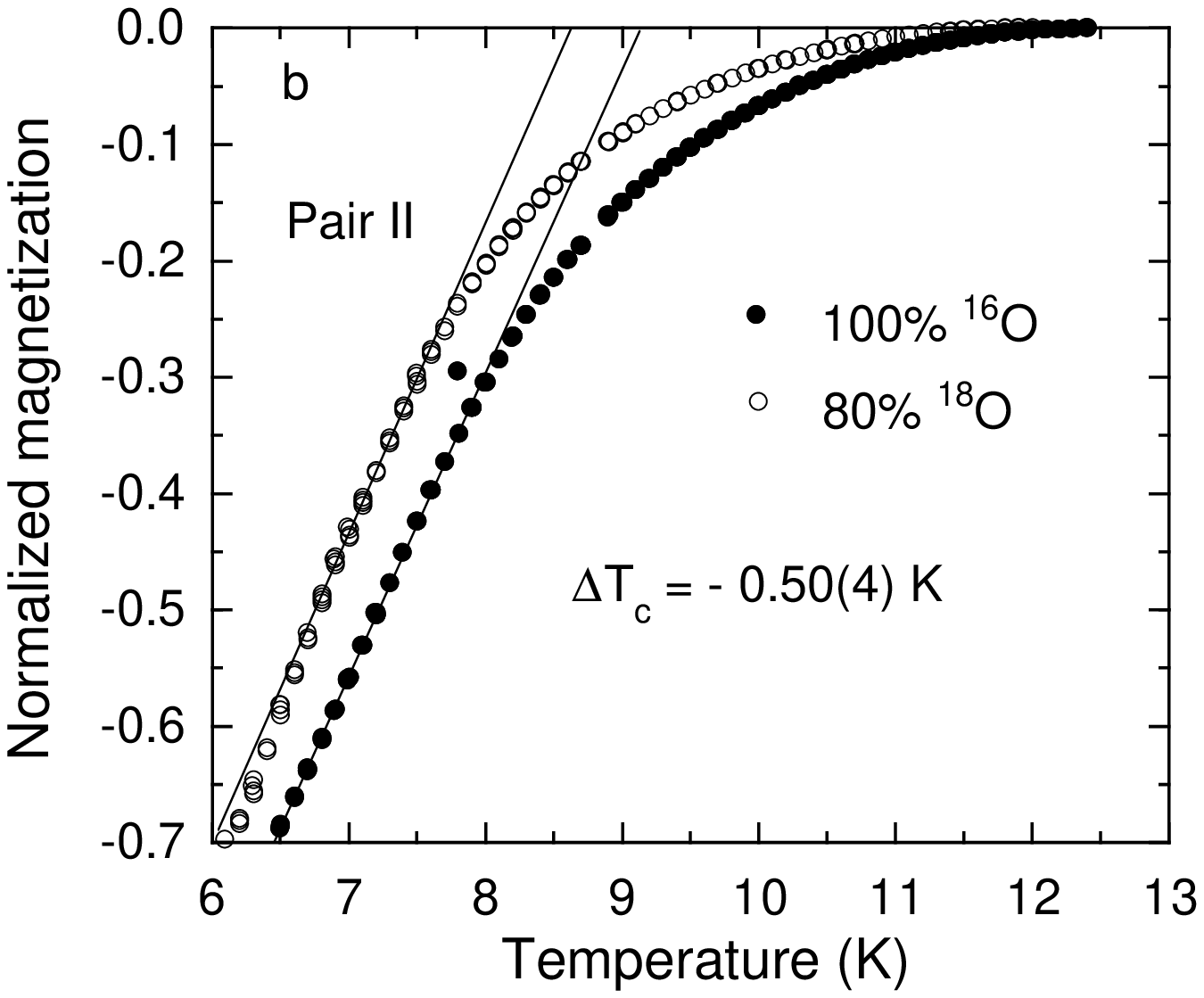}}
	\vspace{1cm}
	\caption[~]{The normalized magnetizations for the
$^{16}$O and $^{18}$O samples of pair I and pair II. The isotope shifts of
$T_{c}$ are determined from the linear portion of the magnetization
data extended to the base line at zero magnetization.}
	\protect\label{Fig.4}
\end{figure}
\newpage
~\\
~\\
\begin{figure}[b]
\vspace{4cm}
    \ForceWidth{14cm}
	\centerline{\BoxedEPSF{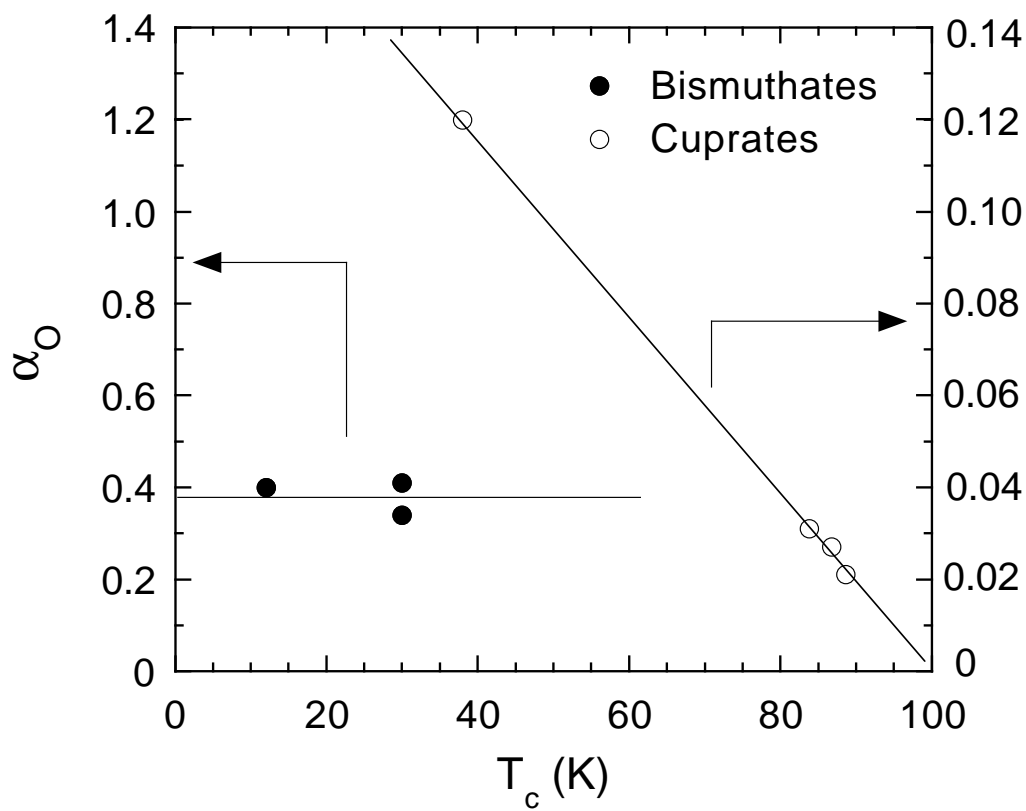}}
	\vspace{1cm}
	\caption[~]{$T_{c}$ dependence of the oxygen-isotope exponent
	$\alpha_{O}$ for optimally-doped bismuthates and cuprates. The data
	for Ba$_{1-x}$K$_{x}$BiO$_{3}$ are from Ref.~\cite{Hinks,Zhao}, and
the
	data for the cuprates are taken from Ref.~\cite{Franck,ZhaoPhysicaC}.
	The data for Sr$_{1-x}$K$_{x}$BiO$_{3}$ ($T_{c}$ = 12 K) are from
this work. }
	\protect\label{Fig.5}
\end{figure}
\end{document}